\documentstyle[11pt]{article}

\newcommand{\be}{\begin{equation}}
\newcommand{\ee}{\end{equation}}
\newcommand{\bea}{\begin{array}}
\newcommand{\ea}{\end{array}}
\newcommand{\beqa}{\begin{eqnarray}}
\newcommand{\eeqa}{\end{eqnarray}}
\newcommand{\bean}{\begin{eqnarray*}}
\newcommand{\eean}{\end{eqnarray*}}

\def\up#1{\leavevmode \raise.16ex\hbox{#1}}

\newcommand{\gapproxeq}{\lower .7ex\hbox{$\;\stackrel{\textstyle >}{\sim}\;$}}
\newcommand{\lapproxeq}{\lower .7ex\hbox{$\;\stackrel{\textstyle <}{\sim}\;$}}

\newcounter{appendice}

\def\thebibliography#1{{\bf REFERENCES\markboth
 {REFERENCES}{REFERENCES}}\list
 {[\arabic{enumi}]}{\settowidth\labelwidth{[#1]}\leftmargin\labelwidth
 \advance\leftmargin\labelsep
 \usecounter{enumi}}
 \def\newblock{\hskip .11em plus .33em minus -.07em}
 \sloppy
 \sfcode`\.=1000\relax}

\begin{document}

\title{\hfill $\mbox{\small{
$\stackrel{\rm\textstyle SU-4240-671\quad}
{\rm\textstyle UAHEP9717 \quad}
$}}$ \\[1truecm]
Properties of Quantum Hall Skyrmions from Anomalies}
\author{S. Baez$^{a}$, A. P. Balachandran$^{a}$, A. Stern$^{b}$ and
A. Travesset $^{a}$}
\maketitle
\thispagestyle{empty}

\begin{center}
{\it a)  Department of Physics, Syracuse University,\\ Syracuse,
New York 13244-1130,  USA\\}
{\it b) Department of Physics, University of Alabama,\\
Tuscaloosa, Alabama 35487, USA}
\end{center}

\begin{abstract}

It is well known that the Fractional Quantum Hall Effect (FQHE) may be
effectively represented by a Chern-Simons  theory.
In order to incorporate QH Skyrmions, we couple this theory to
the topological spin current, and include the Hopf term.
The cancellation
of anomalies for chiral edge states, and the proviso that
Skyrmions may be created
and destroyed at the edge, fixes the coefficients of these new terms.
Consequently, the charge and the spin of the Skyrmion are uniquely
determined.  For those two quantities we find the values $e\nu N_{Sky}$
and  $\nu N_{Sky}/2$, respectively,  where $e$ is electron charge, $\nu$
is the filling fraction and $N_{Sky}$ is the Skyrmion winding number.
We also add  terms to the action so that the classical spin
fluctuations in the bulk satisfy the standard equations of a ferromagnet,
with spin waves that propagate with the classical drift velocity of
the electron.

\end{abstract}

\bigskip
\bigskip

The FQHE admits a Landau-Ginzburg description in terms of
a complex doublet of bosonic fields
$\Psi=\left(\begin{array}{c} \psi_1 \\
\psi_2 \end{array} \right) $ and a statistical
Chern-Simons gauge field \cite{zhkl}.
The Chern-Simons coupling is chosen such that each ``bosonized''
electron carries an odd number of elementary flux units, yielding
fermionic
statistics. The Landau-Ginzburg ground state is given
by $\psi_1=\sqrt{\rho_0}$, $\psi_2=0$, where
 the ground state
density of electrons  $\rho_0$ is equal to the filling fraction $\nu$ times the
external
magnetic field
divided by the elementary flux unit $2\pi/e$.
 The corresponding wave function,
when expressed in position space, is nothing but the Laughlin wave function.

  The lowest lying excitations around
the ground state are described by the quasiparticle and quasihole
Laughlin wave functions.
The presence of a Zeeman interaction naively precludes any dynamics for
the spin degrees of freedom, and
  consequently the quasiparticles and quasiholes are fully polarized.
In the Landau-Ginzburg description, those excitations are
associated with vortices.  More specifically, they are field
configurations  where the
phase of $ \psi_1 $ has a nonzero winding, and
$\psi_2=0$ everywhere.  The magnitude of $\psi_1$,
which is  the square root of the density of electrons,
vanishes at points associated with the origin of the vortices.

On the other hand, it was noticed \cite{Halp} that in  cases where
the gyromagnetic ratio is small, as for example in $GaAs$ samples,
the spin degrees of freedom  play a dynamical
role, and lowest lying excitations above the ground state have some
``ferromagnetic'' properties\cite{Sondhi}. In the Landau-Ginzburg
description, this situation is realized by allowing  for a
nonzero $\psi_2$ in the
spatial domain, which for the moment we define to be all of ${\bf R}^2$.
If we
also assume that the number density  $\Psi^{\dagger}\Psi$ never
vanishes  on ${\bf R}^2$,  it is  possible to everywhere identify an
$SU(2)$
field degree of freedom $g$, associated with spin fluctuations.
 A $U(1)$ subgroup
(which we take to be generated by the third  Pauli matrix $\sigma_3$)
of $SU(2)$ is gauged via the coupling to a statistical gauge
field (the same gauge field mentioned above), so that the gauge invariant
observables are defined on $S^2$.  Energy finiteness generally demands that
$g$ goes to the above $U(1)$ subgroup,
i.e. $g\rightarrow \exp{i\chi\sigma_3}$,
at spatial infinity.
This corresponds to $\Psi$ going to the ground state
value at spatial infinity, and in effect compactifies
${\bf R}^2$ to $S^2$.     Skyrmions\cite{Skyrme},\cite{Sondhi}
associated with $\Pi_2 (S^2)$ result,
the elements of $\Pi_2(S^2)$ being labeled by the winding number
\be
N_{Sky}=\int_{{\bf R}^2} d^2x \; T^{0}(g)=
\frac{i}{4\pi}\int_{{\bf R}^2} d\; {\rm
Tr}\;\sigma_3
g^\dagger dg \;, \label{}\ee where
  $T^0(g)$ is the time component of the topological  current,
\be T^{\mu}(g)=
\frac{i}{4\pi}\epsilon^{\mu\nu\lambda} \; {\rm  Tr}\sigma_3\partial_\nu
g^\dagger \partial_\lambda g \;. \label{topcur}\ee
(The topological current can be defined for vortices as well, only
there it becomes  singular at the zeros of $\Psi^\dagger\Psi$.)

A description dual to the Landau-Ginzburg model was found useful for the
analysis of vortex dynamics \cite{dual}.  It is of interest to
write down a dual description suitable for the analysis of Skyrmion
dynamics as
well \cite{Stone},\cite{seven}. This is one of the purposes of our letter.
In this
regard, we shall argue in favor of including a Hopf term in the
action. We
shall determine its  coefficient, as well as all coefficients, by
requiring the model, including its edge terms to be anomaly free
\cite{anomaly}.
These coefficients then fix Skyrmion properties, such as
charge and spin.  Arguments have been given in the literature which
show that the charge is equal to the winding number times $\nu$ times the
electron charge\cite{Sondhi},
which we shall confirm. Although there is some debate, it is generally
agreed
that the spin should be $1/2$ for Skyrmions with winding number one at
$\nu=1$ \cite{spin}. We shall confirm this result as well.

We first show what are the consequences of  not including a Hopf
term. Our starting point is the bulk action \cite{Fro}
\be\label{Hall_LAG}
  {\cal S}_H=\int_{\Sigma\times R^1} d^3x \;\biggl(
\frac{\sigma_H}{2}\epsilon^{\mu \nu \lambda} A_{\mu}
\partial_{\nu}  A_{\lambda}-  eA_{\mu}{\cal
 T}^{\mu} \biggr) \label{qhlag} \;,
\ee
where $\Sigma$ is the  two dimensional spatial domain of the sample,
$R^1$ accounts for time, and $e{\cal T}^{\mu}$ is the Skyrmion current.
Additional terms  will be added, but for the moment we just
consider  (\ref{qhlag}).
For us, $A_\mu$ is the external electromagnetic field which is not
a dynamical variable. Its variations therefore just
define the bulk current $J_{em}^\mu$ by
$ J^\mu_{em} = -  \frac{\delta {\cal S}_H}{\delta A_{\mu}}  .$

According to (\ref{Hall_LAG}), the bulk electromagnetic current 
$J^{\mu}_{em}$ is,
\be\label{OUR_MOT1}     J^\mu_{em}= -
\frac{\sigma_H}2 \epsilon^{\mu \nu \lambda} F_{\nu\lambda}+
e {\cal T}^{\mu} \; .\label{Halleq}
\ee
For consistency  the current $J^\mu_{em}$, and consequently
 ${\cal  T}^\mu$, must be conserved.  This is the case for
 ${\cal  T}^\mu$
 proportional to the topological current $T^\mu$\cite{gov1},\cite{soto}:
\be {\cal T}^\mu =\kappa
 T^\mu \;.\ee
Eq.  (\ref{Halleq})   implies that the electric charge density is
$ J^{0}_{em}=-\sigma_H F_{12}+e \kappa T^{0} $. Integrating it over
the whole sample gives the total electric charge 
as $-eN_{el}+e\kappa N_{Sky}$, where $N_{el}$ is the total number
of electrons at filling fraction $\nu$ \footnote{Here we have used  
the result $\sigma_H=\frac{e^2\nu}{2\pi}$.}.
Thus $\kappa$ times $e$ is the charge of a Skyrmion of unit winding number.

We now examine under what conditions the bulk action $ {\cal S}_H$ is
consistent with the existence of chiral edge currents.
For the case of filling fraction $\nu=1$, there is a
single edge current on the boundary  $\partial \Sigma$
 of $\Sigma$, which may be represented by a 2d massless chiral
relativistic Dirac fermion \cite{Fro} \cite{soto},
while for fractional values of $\nu$ one gets a Luttinger liquid.
Chirality implies that the electromagnetic current $J_{em}^\mu$ of the
edge fermions  satisfies
$   J_-^{em} =   \frac{1}{\sqrt{2}}(J^0_{em} + J^1_{em})
 =0 $.  In the quantum theory
 this is known to lead to an anomaly, i.e. $ \partial_\mu J^\mu_{em} \ne 0$.

It is convenient to bosonize the edge theory \cite{bal}, and for this
we shall introduce a scalar field $\phi$ on   $\partial \Sigma$.  In terms
of this field,  chirality will mean the following:
\be
 {\cal D}_-\phi= f(x^-) \;,\label{boschi}
\ee
where $x^-={1\over\sqrt{2}}
(x^0-x^1)$, ${\cal D}_\pm={1\over\sqrt{2}}({\cal D}_0\pm{\cal D}_1)$ and
$ {\cal D}_{\mu}$ denotes a covariant derivative.
(Usually the more restrictive condition $f(x^-)=0$ is assumed, but  
(\ref{boschi}) seems enough for us.) 

To proceed  we shall  pose an action principle for the edge field $\phi$.
The edge action  ${\cal S}_{\partial
\Sigma\times R^1}$   should be such that:
i) The total action  ${\cal S}=
{\cal S}_H+{\cal S}_{\partial \Sigma\times R^1}$ is gauge invariant.
ii) It is consistent with chirality, i.e. (\ref{boschi}).
We will show that these two conditions lead to a chiral electromagnetic
 current  $ J_-^{em} =0$, which at the boundary
 is defined by
$ J^\mu_{em} = -  \frac{\delta {\cal S}}{\delta A_{\mu}}|_{\partial
\Sigma \times R^1}$ .  Requirements
i) and ii) also lead to the anomaly.  For this recall  that   the one loop
effects responsible for the anomaly in the
fermionic theory  appear at tree level
 in the bosonized theory.   Thus we can expect  to recover
 the anomaly from the classical equation of motion
 for $\phi$.

We begin by addressing the issue i) of gauge invariance.
If we ignore boundary effects, the bulk action
is separately gauge invariant under
transformations of the electromagnetic potentials $A_\mu$,
\be A_\mu \rightarrow A_\mu + \partial_\mu \Lambda \;,
\label{gtoA}\ee as well as under transformations
 of the
fields $g$,
\be
g \rightarrow
g  \;e^{i \lambda \sigma_3}
\;,
\label{GAUGEM}\ee where both $\Lambda$ and $\lambda$ are functions of
space-time coordinates.
On the other hand,
taking into account   the
 boundary $\partial \Sigma$, one
finds instead that
 (\ref{gtoA})
 gives the surface terms
\be \delta
{\cal S}_H=
-\frac{\sigma_H}{2}\int_{
\partial
\Sigma\times R^1}
 d\Lambda\wedge A +
\frac{e\kappa i}{4\pi}\int_{\partial
\Sigma\times R^1}
 d\Lambda\wedge
 {\rm Tr}\sigma_3g^\dagger dg
\;,\label{blkvar2}
\ee
while gauge invariance under transformations (\ref{GAUGEM}) persists.
We now specify that under gauge transformations
(\ref{gtoA}), the edge field $\phi$
transforms according to
\be \phi \rightarrow \phi + e\Lambda \;.\label{gtpeL}\ee
Then
we can cancel both of the above boundary terms in
 (\ref{blkvar2})
 if we assume the following action for the scalar field
$\phi$:
  \be
{\cal S}_{\partial
\Sigma\times R^1}
 =\frac{R^2}{8\pi}\int_{\partial
\Sigma\times R^1}
 d^2x\;
({\cal D}_\mu\phi)^2
+ \frac{\sigma_H}{2e}\int_{\partial
\Sigma\times R^1}
 d\phi\wedge A
-\frac{\kappa i}{4\pi}\int_{\partial
\Sigma\times R^1}
 d\phi
\wedge
 {\rm Tr}\sigma_3g^\dagger dg
 \;.
\label{stbdac2}
\ee
In (\ref{stbdac2})
 we have added a kinetic energy term for $\phi$,
where the covariant derivative is  defined by
${\cal D}_\mu\phi=
\partial_\mu\phi  -eA_\mu $.  The coefficient $R$ is real and is known to
correspond to the square root of the filling fraction $\nu$.  In this
regard, a straightforward quantization of the edge Lagrangian shows
that $R^2$ is the ratio of  odd integers, and more generally, that entire
hierarchies of filling fractions can be obtained\cite{bal}.

Concerning ii), extremizing (\ref{stbdac2}) with respect to $\phi$ gives
\be\label{aneq2}
R^2 \partial_\mu {\cal  D}^\mu \phi = -\frac{2\pi\sigma_H}{e}F_{01}
-4\pi{\cal T}^r
\;,\ee   $F_{01}$ being the electric
field strength at the boundary and the index $r$   denoting the
direction normal to the surface.  This equation can be
rewritten as
\be\label{aneq20}2
R^2 \partial_+ {\cal  D}_- \phi =(eR^2 -\frac{2\pi\sigma_H}{e})F_{01}
-4\pi{\cal T}^r
\;,\ee using $\partial_+={1\over\sqrt{2}}(\partial_0+\partial_1)$
 and  $diag(1,-1)$ for the Lorentz metric.  But the chirality condition
(\ref{boschi}) requires that the right hand side of
(\ref{aneq20}) vanishes.  For this we can set
\be \sigma_H=\frac{e^2 R^2}{2\pi} \;, \label{Rsqnu2}\ee
 which is the  usual relation for the Hall conductivity (after  identifying
 $R^2$
with the filling fraction $\nu$).  But we also need
\be
{\cal T}^r = 0 \quad {\rm  at }
\;\partial \Sigma \;.\label{tccp}
\ee
From  (\ref{Rsqnu2}) and (\ref{tccp}),  variations of $A_\mu$ give the
following result for the edge current
\be\label{bos_curr_el}
    J^{\mu}_{em} = -  \frac{\delta {\cal S}}{\delta A_{\mu}}|_{\partial
\Sigma \times R^1}=\frac{e   R^2}{4\pi}
\left( {\cal D}^{\mu}+\epsilon^{\mu \nu}{\cal D}_{\nu} \right)\phi\;,\ee
  and thus it is chiral, i.e. $ J_-^{em} =0$.  Here
  $\epsilon^{01}=-\epsilon^{10}=1$.  By taking its divergence we also recover
the anomaly:
\be \partial_\mu J^\mu_{em}  =
    \frac{e   R^2}{4\pi} \partial_\mu
\left( {\cal D}^{\mu}+\epsilon^{\mu \nu}{\cal D}_{\nu} \right) \phi
 =       -   \frac{e^2 R^2}{2\pi} F_{01}
\;, \label{dvj} \ee where we again used (\ref{Rsqnu2}) and (\ref{tccp}).

In order to satisfy chirality in the above discussion,
 we needed not only to constrain the values of coefficients, but we also
found it necessary to impose a boundary condition
 (\ref{tccp}) on the topological current.  As a result, the topological
flux, and moreover Skyrmions, cannot  penetrate
the edge.  Thus, provided $g$ is everywhere defined in $\Sigma$, the total
Skyrmion number within the bulk
$\int_{\Sigma} d^2x \; T^{0}(g)$
is a conserved quantity, and for example, a nonzero value for the total
topological charge
cannot be adiabatically  generated from the ground state.

Below, we generalize to the  situation where the total
Skyrmion number in the bulk is ${\it not}$
 restricted to being a constant.  For this we need to drop
 the boundary  condition (\ref{tccp}), and thus allow for
a nonzero topological flux into or out of the sample.
One may interpret this as Skyrmions
being created or destroyed at the edges.\footnote{Viewing vortices as
 punctures or holes in $\Sigma$,
they too then    act as  sinks and sources of topological flux.}
For this purpose, we consider an extension of the above
description, where  the Hopf term
\be
{\cal S}_{WZ} =\frac{\Theta}{24 \pi^2}
\int_{\Sigma\times R^1}{\rm Tr}( g^{\dagger} dg)^3 \label{WZterm}
\ee                is added
 to the bulk action ${\cal S}_H$.
 [Note that
(\ref{WZterm}) is a local version of the Hopf term]. This term does not affect
the classical equations of motion since it is the integral of a  closed
three form.
It is not well defined for vortices, and must  be suitably
regularized in that case.  On the other hand,
the  Hopf term is well defined for Skyrmions.
Its utility  is in the fact that it provides a direct way for
computing the Skyrmion's intrinsic spin, as we now show. Let
$g^{(N_{Sky},0)}(\vec{x})$   denote a static field
configuration which is nontrivial in a spatial domain ${\cal
V}\subset\Sigma$
and  goes to $\exp{i\chi \sigma_3}$ at the boundary  $\partial {\cal V}$
of this region.
More generally, we can define a one parameter family of configurations,
using
\be
 g^{(N_{Sky},\theta)}(\vec{ x})
 = e^{i\theta \sigma_3 /2} \
g^{(N_{Sky},0)}(\vec{ x}) \ e^{-i\theta \sigma_3 /2} \;,\label{gnt}
\ee
which corresponds to a spin
rotation  by an angle $\theta$. The gauge choice  made in
(\ref{gnt}) is such that the vacuum values (which have the form
$ e^{i \chi \sigma_3}$)
of $g^{(N_{Sky},\theta)}(\vec{ x})$ are invariant under
rotations.\cite{spnispn}
(This fact was neglected in \cite{book}.) As a result of this choice,
 $g^{(N_{Sky},\theta)}$ evaluated at the edge
does not depend on $\theta$, and hence the edge action is unaffected by
such rotations.
Now we consider an adiabatic rotation by
$2\pi$. Thus we set $\theta=\theta(t)$, with
$\theta(-\infty)=0$ and $\theta(\infty)=2\pi$.  To get the spin we
compute the classical bulk action, specifically  $  {\cal S}_{WZ}$,
for this process.  We get
\begin{eqnarray} {\cal  S}_{WZ} = \int^\infty_{-\infty}dt
\int_{\cal V} d^2x \; {\cal  L}_{WZ}& =&\frac{i\Theta}{8\pi}
\int_{\cal V} {\rm Tr}\;\sigma_3 \left[(dg^{(N_{Sky},0)}
{g^{(N_{Sky},0)}}^\dagger)^2-({g^{(N_{Sky},0)}}^{\dagger}dg^{(N_{Sky},0)})^2
\right]\cr
& =&\frac{i\Theta}{8\pi}
\int_{\partial{\cal V}} {\rm Tr}\;\sigma_3 \left[ dg^{(N_{Sky},0)}
{g^{(N_{Sky},0)}}^\dagger+ {g^{(N_{Sky},0)}}^\dagger  dg^{(N_{Sky},0)}
\right]\cr
& =&\frac{i\Theta}{4\pi}
\int_{\partial{\cal V}} {\rm Tr}\;\sigma_3
{g^{(N_{Sky},0)}}^\dagger dg^{(N_{Sky}
,0)} \cr
&=&\Theta\int_{\cal V} d^2x \; T^{0}(g^{(N_{Sky},0)})
 \equiv\Theta N_{Sky}\;,\end{eqnarray}
 where we have used Stoke's theorem.
 Since the action changes by $\Theta $ times the
winding number under a $2\pi$ rotation, its spin (up to an integer) is
\be\frac{\Theta N_{Sky}}{2\pi}\;.\label{spin}\ee

We thus need the numerical value of $\Theta$ to determine the spin.  For
this purpose we now reexamine the boundary dynamics taking into
account the  Hopf term.  We once again require i) gauge invariance and
ii) chirality.

Concerning i),
  as before, the bulk action is not invariant under
  gauge transformations (\ref{gtoA})
  of the electromagnetic potentials $A_\mu$.
In addition,  unlike before, it is  not invariant under gauge
transformations (\ref{GAUGEM})  of the fields $g$.
From ${\cal S}_{WZ}$ we pick up  the surface term
\be \delta{\cal S}_{WZ}=
\frac{i\Theta}{8\pi^2}\int_{\partial\Sigma\times R^1} d\lambda\wedge
 {\rm Tr}\sigma_3g^\dagger dg\;.\label{blkvar3}\ee
To cancel this variation along with  (\ref{blkvar2}), we once again assume
the existence of an edge field $\phi$ which transforms like (\ref{gtpeL}),
simultaneously with the  gauge transformations (\ref{gtoA})
  of the electromagnetic potentials $A_\mu$.
We further specify that $\phi$
 transforms according to
 \be \phi \rightarrow \phi + \lambda
\;,\label{gtsf}\ee simultaneously
 with the gauge transformations (\ref{GAUGEM}) of the fields $g$.
Then we can cancel both of the  boundary terms  (\ref{blkvar2}) and
(\ref{blkvar3}), making our theory anomaly free,
if we assume the following action for the scalar field $\phi$:
  \begin{eqnarray}\label{stbdac5}
{\cal S}_{\partial\Sigma\times R^1}
 &=&\frac{R^2}{8\pi}\int_{\partial\Sigma\times R^1}
 d^2x\;({\cal D}_\mu\phi)^2+ \frac{\sigma_H}{2e}
\int_{\partial\Sigma\times R^1}
 d\phi\wedge A \\
& &- \frac{i}{4}\int_{\partial\Sigma\times R^1}
\biggl(\frac{\Theta}{2\pi^2} d\phi
+\frac{\sigma_H}{e} A \biggr)\wedge {\rm Tr}\sigma_3 g^\dagger dg
\nonumber \;,
\end{eqnarray}
provided we also impose that \be
\kappa  = \frac{\pi\sigma_H}{e^2}+\frac{\Theta}{2\pi}
 \;.\label{formfk}\ee   Since
$\phi$ admits   gauge transformations  (\ref{gtsf}), as well as
 (\ref{gtpeL}), we must
redefine the covariant derivative appearing in (\ref{stbdac5}) according
to    \be    {\cal D}_\mu\phi=
\partial_\mu\phi  -\beta_\mu\;,\qquad \beta_\mu  =
eA_\mu -\frac{i}{2}{\rm Tr}\sigma_3 g^\dagger
\partial_\mu g\;.\ee

With regard  to ii),
 the equation of motion for $\phi$ is  \be
R^2 \partial_\mu {\cal  D}^\mu \phi = -\frac{2\pi\sigma_H}{e}F_{01}-
2{\Theta} T^r\;, \ee
which can be rewritten as
\be\label{aneq21}2
R^2 \partial_+ {\cal  D}_- \phi =(eR^2 -\frac{2\pi\sigma_H}{e})F_{01}
+2(\pi R^2-\Theta){\cal T}^r
\;.\ee
  We recover the chirality condition
(\ref{boschi}) upon setting
\be\Theta
=\pi R^2\;,\label{Rsqnu5}
\ee as well as (\ref{Rsqnu2}).
From  (\ref{Rsqnu2}) and (\ref{Rsqnu5}),  variations of $A_\mu$ again give the
 the edge current in the form of
(\ref{bos_curr_el}) (although the covariant derivative is now
 defined differently)
 and thus  $ J_-^{em} =0$.   By taking its divergence we get
the anomaly equation:
\be \partial_\mu J^\mu_{em} =-\frac{eR^2}{2\pi}
\epsilon^{\mu\nu} \partial_\mu\beta_\nu
\;. \ee

  Thus now we can satisfy the criterion of chirality
 without imposing any boundary conditions on the topological current.
     Substituting (\ref{Rsqnu5})  into
 (\ref{formfk}) (and using $ R^2=\nu$)
  also fixes $\kappa$ to be the filling fraction.  It follows that the
Skyrmion charge is $e\nu N_{Sky}$.
 Eqs.  (\ref{spin}) and (\ref{Rsqnu5})
then give the value for the spin to be $ \frac{N_{Sky}
\nu}2 $.  Therefore, within the above assumptions, a winding
number one Skyrmion is a fermion when the filling fraction is one.

We note further that using the above values for the constants, we can
simplify
the bulk action
 $ {\cal S}_H + {\cal S}_{WZ}$
 to the single Chern-Simons term
\be \frac\nu{4\pi} \int
_{\Sigma \times R^1}
\beta \wedge    d\beta   \ee
plus the surface term
\be \frac{ie\nu}{8\pi}\int
_{\partial\Sigma \times R^1}A\wedge {\rm Tr} \sigma_3 g^\dagger dg \;.\ee
This surface term cancels the last term in (\ref{stbdac5}), and
consequently the boundary action simplifies to
\be
 \frac{\nu}{8\pi}\int_{\partial\Sigma\times R^1}
 d^2x\;(\partial_\mu\phi-\beta_\mu)^2+
\frac{\nu}{4\pi}\int_{\partial\Sigma\times R^1}
 d\phi\wedge \beta \;.\ee
Thus if no additional terms are present, and  there are
no vortex singularities present in the sample, the bulk plus edge action
can be written entirely
in terms of the redefined Chern-Simons connection $\beta_\mu$ and the edge
field $\phi$.

In the above treatments we have just considered
the topological sector of our dual description. We
have not incorporated the terms responsible for spin fluctuations
in the bulk, and other possible higher derivative terms. 
It is important to have in mind that there are spin
waves, the Goldstone modes associated with a ferromagnetic ground
state.  We will include those fluctuations
by demanding that the classical equations of motion are those
of a ferromagnet, and have Skyrmions among its solutions.
That is achieved by including  a kinematic term for the Skyrmion:
\be {\cal S}_S = -  \frac{\eta}{2}\int_{\Sigma \times R^1}d^3 x
 (\partial_i  n_a)^2  \;,\quad a=1,2,3 \;,\ee
where $n_a$ is a unit vector field  defined by
$n_a \sigma_a= g\sigma_3 g^{\dagger}$ and $\eta$ is a constant.
Now, the dynamics in the bulk cannot be described solely by the
Chern-Simons connection $\beta_\mu$.  Starting from the total bulk
 action    $  {\cal S}_H + {\cal S}_{WZ} + {\cal S}_S$,
 the dynamics for the spin degrees of freedom is readily
obtained
from infinitesimal variations $\delta g = i\epsilon_a \sigma_a g$
of $g$:
\be
 2\eta
 (\nabla^2 n \times n)_a
=\frac{e\nu}{2\pi}\epsilon^{\mu\nu\lambda}
\partial_\mu n_a \partial_\nu
A_\lambda \;.\ee
which, we can rewrite as,
\be\label{EQ_MOT2}
\partial_0n_a-v_j\partial_jn_a-\frac{2\eta}{\rho_0}
(\nabla^2 n \times n)_a
 =0\; ,\label{lleq}
\ee
where $\rho_0$ is the ground state density of electrons, $\vec{v}$ 
is their classical drift velocity ( which is
in the plane of the sample and perpendicular to the electric field $E$,
with total velocity $|v|=cE/B$), and the first two
terms denote the convective derivative of $n_a$. This is the same equation
that follows from the Landau-Ginzburg formalism \cite{Stone} as well.

\bigskip

\bigskip
{\bf Acknowledgement}
We wish to thank T. R. Govindarajan, R. Shankar and  S. Vaidya
for very helpful discussions.
A. P. B. and A. T. were supported during the course of this work
by the Department of Energy, USA, under contract number
DE-FG02-85ERR40231.
A. S. was supported during the course of this work
by the Department of Energy, USA, under contract number
DE-FG05-84ER40141.

\end{document}